# EXTRACTION OF THE PROTON BEAM FROM THE 70 GEV IHEP ACCELERATOR

A.G. Afonin\*, V.T. Baranov\*, V.M. Biryukov\*, Yu. M. Ivanov\*\*, A.A. Kardash\*, V.I. Kotov\*,
V.A. Maisheev\*, V.I. Terekhov\*, E.F. Troyanov\*, Yu.S. Fedotov\*, V.N. Chepegin\*, Yu.A. Chesnokov\*

\* IHEP, Protvino, Moscow oblast, 142284 Russia
\*\* INPh, Gatchina, St. Petersburg, 188350 Russia

*Abstract*

This paper presents the physical background for particle extraction from IHEP accelerator using short bent silicon crystals, analyses the results of the studies, considers in detail the regime of simultaneous work of crystal extraction and several internal targets. It is experimentally shown that the use of short crystals allows the extraction of beams with intensity of $\sim 10^{12}$ proton/cycle with efficiency of $\sim 85\%$.

## INTRODUCTION

Over a period of almost thirty years, several laboratories around the world have developed a new method of steering high-energy particle beams, based on the use of bent crystals. Employing this technique, the first extracted charged particle beams were obtained with a moderate intensity of up to $10^7$ particles per pulse at extraction efficiencies of $10^{-4}$-$10^{-3}$. In 1989, at the IHEP U-70 accelerator, a proton beam with energies of 50 and 70 Gev was extracted into one of the existing channels. To direct the extracted beam into this channel, it was necessary to bent a (111) silicon crystal 65*15*0.6 mm in size to a large angle, ~85 mrad. To increase the extraction efficiency, one has to employ short crystals bent at small angle. In this case, as calculations for the IHEP accelerator showed, crystals with a 1-2 mm length and bending angles of 1-2 mrad can provide particle extraction with an efficiency of ~85%. The development of such an extraction for the IHEP accelerator was started in 1997 and mostly completed by the end of 2001.

## EXTRACTION SCHEMES AT THE IHEP PROTON ACCELERATOR

The moderate extraction efficiency in experiments with long crystals was caused by the fact that particles are captured into the channeling regime mostly during the first passage through the crystal. Particles not captured into the channeling regime are strongly scattered and are eventually lost. To attain efficient multipass extraction, when particles not captured during the first passage can be captured during subsequent passages, short-length crystals with small bending angles are required. In this circumstance, the efficiency of particle capture into the channeling regime can be estimated from the following considerations: as a rule, a regular structure of the crystal surface layer 10-30 μm thick is disturbed during its treatment. Particles passing through this layer will be subject to multiple Coulomb scattering, and a fraction of the beam is lost as a result of nuclear interactions. At the next encounter with the crystal, this time in its regular region, the beam at the entrance face will have the distribution with the root-means-square angle given by

$$\sigma = \frac{13,6(M \ni B/c)}{p(M \ni B/c)} \sqrt{\frac{L}{L_R}} \left[1 + \frac{1}{9}\lg\left(\frac{L}{L_R}\right)\right], \quad (1)$$

where $L$ is the crystal length and, $L_R$ is the radiation length equal to 9.36 sm of silicon. During subsequent passages through the crystal, the root-means-square scattering angle $\sigma$ will increase when $\sigma_k = \sqrt{k}\sigma$, where $k$ is the number of passages.

During the first passage through the regular crystal region, the fraction of extracted particles is $P_1 = F_1 W$, where $W = e^{-L/L_N}$ defines the fraction of particles remaining in the beam after passing through the amorphous layer, and $L_N$ is the nuclear length equal to 45 sm of silicon. The quantity $F_1$ is the efficiency of particle deflection at a set angle $\Theta$. In the case of the Gaussian distribution of particles at the crystal entrance face, takes on the form

$$F_1 = \Phi(x) \frac{\pi x_C}{2d} (1-\rho)^2 \exp\left(-\frac{\Theta}{\rho(1-\rho)^2 \Theta_D}\right). \quad (2)$$

Where $\Phi(x) = \frac{2}{\sqrt{2\pi}} \int_0^x e^{-t^2/2} dt$ is the probability integral, $\rho = \frac{R_C}{R}$, $x = \frac{\Psi_C}{\sigma}$ ($x = \frac{\Psi_C}{\sqrt{k}\sigma}$ for the $k$th passage). After the first passage $W(1-F_1)$ particles will not be captured. During the second passage, $W^2(1-F_1)F_2$ particles will be removed from the fraction of particles not captured into the channeling regime, and $W^2(1-F_1)(1-F_2)$ particles will remain. During the third passage, the fraction $W^3(1-F_1)(1-F_2)F_3$ of remaining particles will be captured into the channeling regime, and so on. This sequence of the capture probabilities during each passage can be written as

$$P = \sum_{K=1}^{N} P_K = F_1 W + \sum_{K=2}^{N} F_K W^K \prod_{m=1}^{K-1}(1-F_m). \quad (3)$$

The proton extraction efficiency calculated using formulas (2) and (3) is shown in Fig.1. These calculations carried out for (110) silicon crystals of various lengths, deflecting a 70-Gev proton beam at a fixed angle of 2 mrad.

The results of direct numerical simulation (closed squares) are also shown there. The difference in the results for short crystal lengths is due to the fact that the future of the dechanneled particles is disregarded in the calculations using formula (3) in contrast to numerical simulation. However, a certain fraction of dechanneled particles that passed through most of the crystal length in the channeling mode can also be removed.

We can see in Fig.1 that multipass extraction yields threefold extraction efficiency in comparison with one-pass extraction. In this case, short crystals ~1-2 mm long, provide the highest efficiency (~80%). As for restrictions on crystal thickness, they are rather vague. As estimations show, crystals 0.3-0.6 mm thick can be used without a significant influence on efficiency.

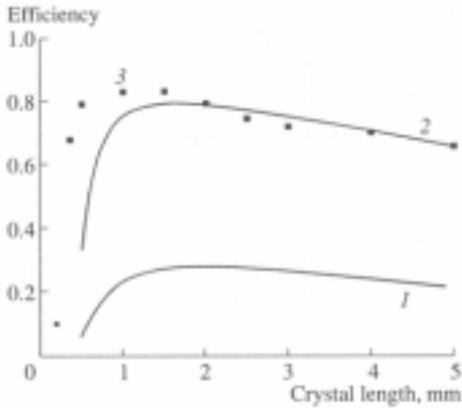

Figure 1: Extraction efficiency vs. crystal length (1) one-pass and (2) multi-pass extraction; squares (3) correspond to direct numerical simulation.

As is known, in cyclic strong-focusing accelerator particle motion with respect to the equilibrium orbit in vertical and horizontal planes can be written as

$$x(s) = \sqrt{E\beta}\cos(\Psi + \delta), \quad (4)$$

$$x'(s) = -\sqrt{\frac{E}{\beta}}\alpha\cos(\Psi + \delta) - \sqrt{\frac{E}{\beta}}\sin(\Psi + \delta),$$

where $s$ -is the longitudinal coordinate along the equilibrium orbit; $\alpha$ and $\beta$ are the structure functions of the accelerator; and $\beta' = -2\alpha$, $\Psi' = \frac{1}{\beta}$, $E$ and $\delta$ are arbitrary constants depending on the initial condition.

By eliminating the trigonometric functions we can write

$$\left(x_{\max} = \sqrt{\beta E},\ x' = -\frac{\alpha}{\beta}x_{\max}\right),\ \left(x = -\frac{\alpha}{\gamma}x'_{\max},\ x'_{\max} = \sqrt{\gamma E}\right). \quad (5)$$

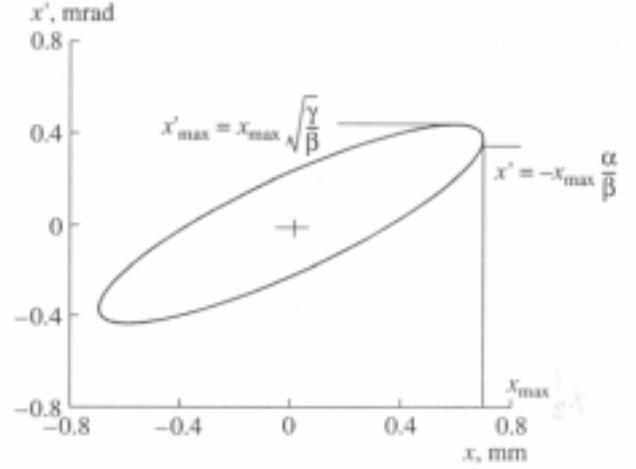

Figure 2: Phase diagram of the circulating beam.

One can see in Fig. 2 that $x_{\max}$ and $x'_{\max}$ play the role of spatial and angular beam envelopes. If the beam half-size $x_{\max}$ is measured, Eq. (5) allows the calculation of the phase area $S = \pi E = \pi \frac{x_{\max}^2}{\beta}$ and $x'_{\max} = x_{\max}\sqrt{\frac{\gamma}{\beta}}$ of a beam. The phase area of the accelerated beam of the U-70 IHEP accelerator in the horizontal plane is approximately equal $2\pi$ mm mrad at an intensity above $10^{12}$ protons per cycle.

The particle beam, when guided to the crystal, comes into contact with it at the point $x_{\max}$, $x' = -\frac{\alpha}{\beta}x_{\max}$. Since the crystal structure in the surface layer is, as a rule, disturbed, the beam will be subject to multiple scattering during the first passage through the crystal, which results in an increase in the oscillation amplitude,

$$X = \sqrt{x_{\max}^2 + \beta^2\sigma^2},\quad \Delta X = \sqrt{x_{\max}^2 + \beta^2\sigma^2} - x_{\max} = \frac{\beta^2\sigma^2}{2X},$$

where $X \approx x_{\max}$, $\sigma$ is the root-mean-square scattering angle defined by formula (1).

As we have already seen, it is important to take this factor into account when calculating the beam extraction efficiency using a bent crystal.

We will enlarge on two additional circumstances, which should be considered in the development of beam extraction using bent crystals: As we saw, to attain efficient extraction, the deflection angle Θ of the particles caused by the crystal should be small. Therefore, beam extraction from the accelerator requires the installation of additional deflecting magnets at the deflected beam trajectory. The relative positions of the crystal and the first magnet should be such that the extracted beam arrives at the magnet operating aperture.

The second circumstance is associated with the steering of the beam onto the crystal.

To consider these problems, we invoke the known formula defining particle motion in an arbitrary interval $(s_2 - s_1)$ of the accelerator's magnetic structure.

$$\begin{pmatrix} x_2 \\ x_2' \end{pmatrix} = \begin{bmatrix} \sqrt{\frac{\beta_2}{\beta_1}}(\cos\Delta\Psi + \alpha_1 \sin\Delta\Psi) & \sqrt{\beta_1\beta_2}\sin\Delta\Psi \\ -\frac{(1+\alpha_1\alpha_2)\sin\Delta\Psi + (\alpha_2-\alpha_1)\cos\Delta\Psi}{\sqrt{\beta_1\beta_2}} & \sqrt{\frac{\beta_1}{\beta_2}}(\cos\Delta\Psi - \alpha_2 \sin\Delta\Psi) \end{bmatrix} \begin{pmatrix} x_1 \\ x_1' \end{pmatrix}, \quad (6)$$

where $\Delta\Psi = (\Psi_2 - \Psi_1)$ is the phase increment in this interval. To determine the deflection of the beam by the crystal at the position of the first additional magnet, instead of $x_1$ and $x_1'$ in Eq.(6), we substitute their values for $x_{max}$ and $\left(-\frac{\alpha_1}{\beta_1}x_{max} + \Theta\right)$, respectively. After completing some simple algebra, we derive

$$x_2 = x_{max}\sqrt{\frac{\beta_2}{\beta_1}}\cos\Delta\Psi + \sqrt{\beta_1\beta_2}\Theta\sin\Delta\Psi. \quad (7)$$

To optimize the mutual positions of the crystal and the deflecting magnet, we will determine, by solving equation $\frac{dx_2}{d(\Delta\Psi)} = 0$, the optimum phase increment and, from Eq.(7), the optimum value of $x_2$ corresponding to it,

$$(x_2)_{opt} = \sqrt{\frac{\beta_2}{\beta_1}(x_{max}^2 + \beta_1^2\Theta^2)}. \quad (8)$$

The accelerated beam can be steered onto the crystal, e.g., by controlled local distortion of the closed orbit (the so-called bump). To this end, in the simplest case, it is required that two magnetic units of the accelerator are used, placed at a distance of half a wavelength of betatron oscillations from each other. Using additional coils, a magnetic field perturbation ΔB is induced in the first magnetic unit, which increases during beam guidance. As a rule, the magnetic unit length in strong-focusing accelerator is much shorter than the wavelengths of betatron oscillations, and the effect of the perturbing field ΔB is reduced to a change in the orbit angle at the magnet's center $\Delta x_0' = \frac{\Delta B}{B_0}\frac{L}{R_0}$, in the point approximation, where L is the unit length and $B_0$ is the inductance of the main magnetic field in the orbit of radius $R_0$. By substituting the initial values $x_1 = 0$ and $x_1' = \Delta x_0'$, at the center of the perturbing magnetic unit into Eq. (6), for the perturbed orbit, we derive

$$\begin{cases} x_\delta = \sqrt{\beta_1\beta_2}\sin\Delta\Psi \cdot \Delta x_0', \\ x_\delta' = \sqrt{\frac{\beta_1}{\beta_2}}(\cos\Delta\Psi - \alpha_2\sin\Delta\Psi)\Delta x_0' = \frac{x_\delta}{\beta_2}\left(\frac{\cos\Delta\Psi}{\sin\Delta\Psi} - \alpha_2\right) \end{cases} \quad (9)$$

where $\beta_1$ is the structure function at the center of the first magnet, $\beta_2$ and $\alpha_2$ give the crystal's position, and $\Delta\Psi$ is the phase increment between the center of the first magnet and the crystal.

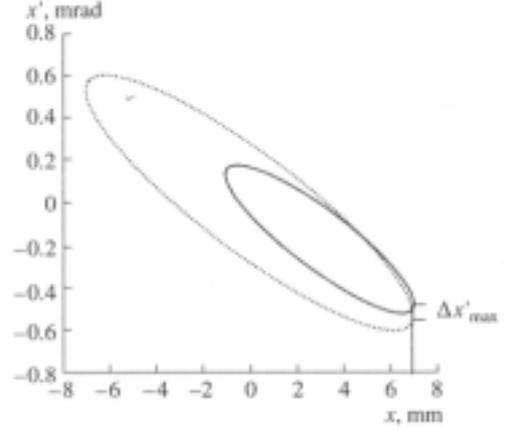

Figure 3: Ellipses Xmax1=7mm and Xmax2=4mm (displaced by the bump by 3 mm).

It follows from Eq.(9) that as the bump increases, its guidance angle to crystal proportionally varies. As a result, an increasing mismatch arises between the angular position of the crystal and the beam (see Fig.3), which eventually decreases the extraction efficiency

$$\frac{\cos\Delta\Psi}{\sin\Delta\Psi} - \alpha_2 = 0. \quad (10)$$

For a canonical ellipse ($\alpha_2 = 0$), this condition is reduced to $\cos\Delta\Psi=0$, i.e., to $\Delta\Psi = \pm\frac{\pi}{2}$ .

In an operating accelerator, these conditions are difficult to meet, since the position where crystals would be installed are usually occupied by other system. In the simplest example, the mismatch in the angle, when the beam is completely guided to the crystal, is given by

$$\Delta x_{max}' = \frac{x_{max}}{\beta_2}\frac{\cos\Delta\Psi}{\sin\Delta\Psi}, \quad (11)$$

where $x_{max}$ is half the beam size. In order that the extraction efficiency is unchanged during beam guidance, $\Delta x_{max}'$ should not exceed the Lindhard angle.

As the accelerated beam is deflected by the crystal at small angles (1-2 mrad), its direct extraction from the accelerator is impossible due to the short straight sections. Therefore, it was also necessary to use several septum magnets from the existing extraction system.

Three single-type stations with bent crystals $S_{i19}$, $S_{i22}$ and $S_{i106}$, were installed in various places in the accelerator (see Fig.4). The subscript means the straight section index or the index of a magnet in which the crystal is installed. Each station can provide proton beam extraction toward an existing extraction channel at a corresponding of bumps.

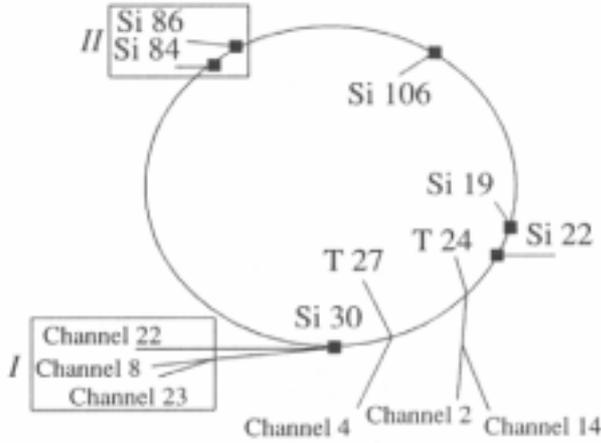

Figure 4: Schematic diagram of beam extraction from the U-70: $S_{i19}$, $S_{i22}$, $S_{i30}$, $S_{i84}$, and $S_{i106}$ are bent crystals; T 24, T 27 are the internal targets; I is the region of experimental systems and II is the region of crystal studies.

Apart from these station $S_{i30}$, $S_{i84}$ and $S_{i86}$ were also placed in the accelerator. Crystalline station $S_{i30}$ is intended to remove a small fraction of particles (~$10^7$ protons) extracted toward beam channel no.8 and to deflect them into channel no.22 (the deflection angle is ~9 mrad). Crystalline stations $S_{i84}$ and $S_{i86}$ are placed in the test region and are intended to test crystals before their installation in the workstations $S_{i19}$, $S_{i22}$ and $S_{i106}$, as well as to carry out studies into, in particular beam collimation modes using bent crystals. Let us enlarge on the chosen schemes of accelerator beam extraction using bent crystals. Figure 5 shows two of these extraction schemes. In one scheme, crystalline station $S_{i19}$, which has two crystals (one is a reserve), is installed in straight section no.19. It provides independent movement of each crystal along the radius, as well as changing its angular orientation with respect to the beam. The beam is guided to the crystal by a local distortion of the orbit using two pairs of magnetic units. Particles guided to the crystal and captured into the channeling regime are deflected by the crystal at an angle of ~1.7-2.5 mrad (depending on the crystal bending angle) and arrive at the aperture of septum magnet SM20, avoiding a septum. After being deflected by the septum magnets SM22 and SM26 (curve 2 in Fig.5), particles are removed from the accelerator vacuum chamber in the 30th straight section. This scheme is attractive because simultaneous beam extraction using the crystal and two internal targets T24 and T27 (curve 1 in Fig.5) is successfully carried out. Using another station $S_{i22}$, placed at the center of magnetic unit no.22, protons were extracted through the septum magnets SM24 and SM26 (curve 3 in Fig.5). In this case, the beam was guided using another two pairs of magnetic unit. When operating with the third station $S_{i106}$ installed in straight section no.106 (not shown in Fig.5), the beam was also guided to the crystal using two pair of magnetic units: one in the feedback mode and another in the dc mode. The beam was extracted through the septum magnets SM20, 22 and 26 and the septum magnets SM24 and SM26. Thus, the third scheme, in the case of faults in the first two, can replace either of them.

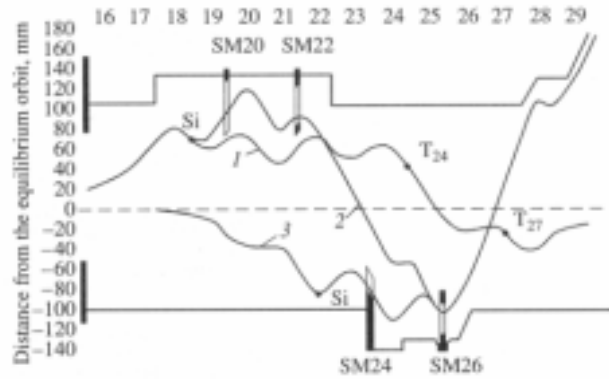

Figure 5: Schematic diagrams of proton beam extraction using bent crystals: (1) the trajectory of circulating beam during its simultaneous guidance onto the crystal in straight section no.19 (Si) and internal target T24 and T27, (2) the trajectory of beam extraction from crystal SS-19 (Si), and (3) the trajectory of beam extraction from crystal Unit 22 (Si).

Table 1 lists the values of the structural functions α and β of the U-70 accelerator for positions of the crystals and the first septum magnets, as well as the phase increments between them. Table 1 also lists the distances $x_2$ (calculated using formula (7)) between the extracted and circulating beams for a crystal with the bending angle $\Theta=2$ mrad and the accelerated beam half-size $x_{max}=7$ mm at the entrance of the first septum magnets.

We can see from Table 1 that the beam separation is sufficient to inject the extracted beam into the aperture of the first septum magnets. Moreover, when extracting the beam according to schemes Si106-24-26 or 20-22-26, as well as Si22-24-26, crystals with bending angles of ~1 mrad can readily be used and (as we shall see below) high extraction efficiencies can be achieved.

Table 1.

| Crystal positions | Crystal | | Position of septum magnets | At the SM positions | | | Distance $x_2$ between the extracted and circulating beam, mm | Septum thickness, mm |
|---|---|---|---|---|---|---|---|---|
| | $\beta_1$, m | $\alpha_1$ | | $\beta_2$, m | $\alpha_2$ | $\Delta\Psi$ | | |
| SS-19 | 24,54 | 1,872 | SS-20 | 21,75 | -1,355 | 0,65 | 28,0 | 7,0 |
| SS-106 | 34,06 | -1,923 | SS-20 | 21,75 | -1,355 | 17,307 | 54,0 | 7,0 |
| Unit 22 | 41,25 | 0,0 | SS-24 | 23,80 | -1,123 | 0,74 | 42,2 | 2,4 |
| SS-106 | 34,06 | -1,923 | SS-24 | 23,80 | -1,123 | 19,336 | 31,8 | 2,4 |

Table 2 lists the structural functions at the centers of the first magnetic units of the system of beam guidance to the crystal and in the crystal positions, as well as the phase increments between them. Table 2 also lists the mismatches $\Delta x'_{max}$ (calculated using formula (11) for $x_{max}$=7 mm) in the angle between the beam and the crystal when the beam is completely guided to the crystal.

Table 2.

| Crystal position | Bump | 1st unit of the bump | | Crystal | | $\Delta\Psi$ | Angular mismatch $\Delta x'_{max}$, μrad |
|---|---|---|---|---|---|---|---|
| | | $\beta_1$, m | $\alpha_1$ | $\beta_2$, m | $\alpha_2$ | | |
| SS-19 | 16-22 | 41,22 | 0,0 | 24,54 | 1,872 | 1,15 | 128 |
| SS-106 | 104-106 | 41,22 | 0,0 | 34,06 | -1,923 | 1,056 | 116 |
| Unit 22 | 20-26 | 36,03 | 0,0 | 41,25 | 0,0 | 1,108 | 85 |

## 3. THE CRYSTAL TYPES USED.

To implement the above-considered schemes of proton beam extraction, it was necessary to solve the very complex problem of bending short silicon crystals at small angles. At the Institute of Nuclear Physics, an O-shaped crystal was developed, which consisted of two U-shaped crystals joined by legs (Fig.6). This technique has been employed at the first stage of our experiments. In this design, the leg thickness is 1 mm. The crystal working region (1) 5*5*0.6 mm$^3$ (length along the beam, height, and thickness) was bent at a small region (2) using a special device.

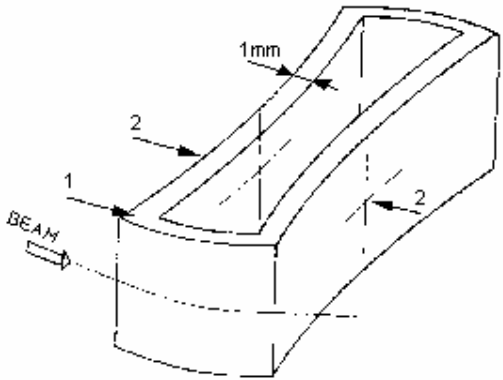

Figure 6: A schematic representation of the O-crystal (1) working region and (2) compressing device position.

A further increase in the extraction efficiency is required to eliminate leg material in the beam path. To this end, a new method for bending a crystal shaped in the form of a narrow strip ~2 mm in length along the beam and 30-40 mm in height (S-crystal) has been developed. The method is based on the anisotropic properties of lattices, (Fig.7).

Specifically, such strip-shaped crystals, ~2 mm in length along the beam and subjected to special chemical polishing, showed the best beam extraction efficiency: ~85%. A general view of the bending device, with a strip-shaped crystal, is shown in Fig. 8.

To apply bent crystals as devises for beam extraction from the accelerator, so-called crystalline deflector stations, and their control system, were developed. Currently, eight stations are installed at the accelerator.

A typical station has two independent mechanisms to move two crystals in the horizontal plane. The precision of crystals positioning with respect to the equilibrium orbit is ~0.1 mm. The crystal is positioned at an angle with respect to the beam guided to the crystal using angular displacement mechanism based on electric step motors. The angular positioning precision is ~13.5 μrad.

Crystalline stations themselves, after being mounted in the accelerator, become a part of vacuum volume of the accelerator. Therefore, their structure is designed for an operating vacuum of 10$^{-7}$ Torr.

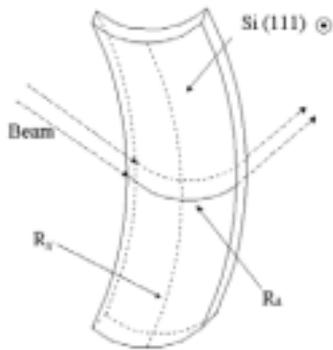

Figure 7: The principle of strip crystal bending.

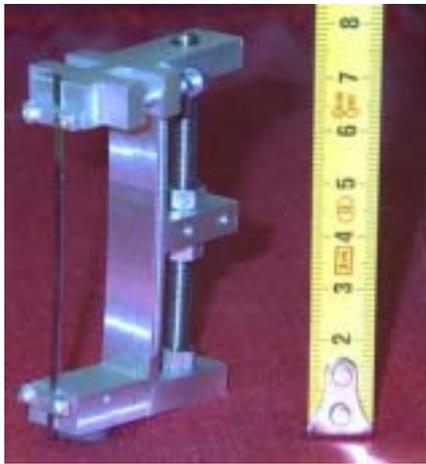

Figure 8: The implemented structure for bending crystal strips.

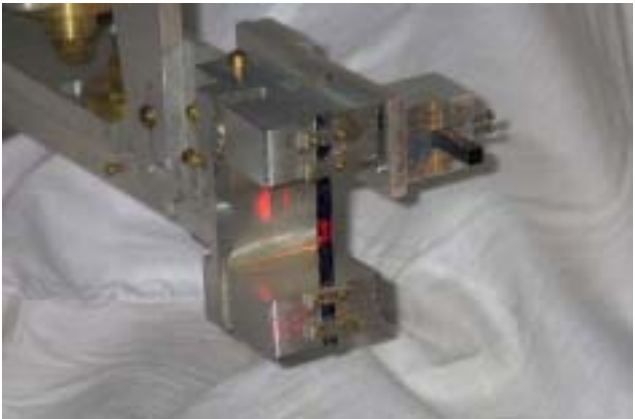

Figure 9: Bent crystals installed in the station.

Figure 9 shows two crystalline deflectors installed in the station: an S-crystal, 5*35*0.5 mm$^3$ (length, height, thickness) in size (left), and O-crystal, 5*5*0.5 mm$^3$ in size (right).

## 4. DIAGNOSTICS OF THE BEAM EXTRACTED BY A BENT CRYSTAL.

The controlled parameters of the extracted beam are its intensity, measured within $10^9$-$10^{12}$ protons per spill, the extraction efficiency, the spill time structure; the beam geometry, at most interesting points of the extraction path, and the radiation loss during extraction. The latter characteristic is not a direct beam parameter, however, it is important for optimizing the steering conditions and also in various studies. Figure 10 shows the layout of the beam diagnostic devices when operating with the crystals installed in straight sections no.106 and 19 of the U-70 accelerator.

When operating with these crystals, beam extraction path are close to the path of slow resonant extraction, which allows the use of a common set of diagnostic devices. At the same time, a number of features characteristic of the channeled beam required special approaches to the measurement technique and procedure. Decreased lateral sizes of the deflected beam require profilometers with the best discreteness. When a low-intensity beam (smaller that $10^7$ protons) is split by the crystal from a high-intensity one, beams whose intensities differ by several orders of magnitude simultaneously pass in two nearby transport channels. When this happens, the low-intensity beam is monitored in the presence of significant background fields induced by the high-intensity beam.

The intensity during the beam spill time is multiply measured. This allows measurement of the spill time structure in a frequency band up to 600 Hz. An acceptable measure accuracy of the total intensity of the extracted beam can be attained only after SEC calibration by a more precise instrument in the same beam.

The secondary emission chamber (SEC30) is intended to measure the intensity of the beam extracted by the crystal. To measure the intensity of the circulating beam steered onto the crystal, and to determine the extraction efficiency, a circulating beam transformer (CBT) is used. The extracted beam transformer (EBT), is placed near the SEC30 to measure the beam intensity during fast extraction. It is used to calibrate the SEC30.

The extraction efficiency is defined as the ratio of the extracted beam intensity, measured using the SEC, to that of the intensity of the beam guided to the crystal, measured using the CBT. The accuracy of the latter is about 1%; thus, the accuracy of the extraction efficiency calculation is defined only by the SEC error and does not exceed 4% in range of $10^9$-$10^{12}$ protons per spill.

An advantage of the SEC is its linear response in a wide range of beam intensities, as well as its high-speed performance. To increase the sensitivity, a combination of several emitters is used.

For slow extraction, such a chamber for measuring the beam intensity is mounted in front of the accelerator exit window (see Fig.10), where a pressure of $10^{-6}$ Torr, required for normal operation of the device, is maintained. Thin (10 μm) Kapton films, with aluminum layers 0.3 μm thick and deposited on both sides, are used as emitters and collectors. The service life of this chamber was experimentally estimated as corresponding to flux of about $10^{19}$ protons/cm2.

The geometric sizes of the extracted beam, i.e., its profile, are determined using multielectrode detectors (profilometers) installed in the beam path.

The layout and main characteristics of profilometers are listed in Table 3. The column "Extraction conditions" contains the following abbreviations: FE is fast extraction, SRE is slow resonant extraction, and CSE is slow extraction using crystals.

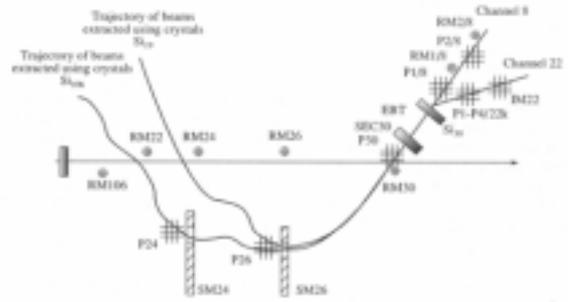

Figure 10: Layout beam diagnostic tools in the extraction path.

Profilometers P24, P26 and P30 installed at the septum magnet entrances and at the ring exit, as well as a number of profilometers in the head regions of transport channels no.8 and no.22, monitor the beam geometry characteristics in both planes.

Figure 11 shows the capillary profilometer (P26) with a drive based on a small-size solenoid. Most of the monitors are equipped with drivers and can be removed from the beam.

Table 3

| Layout or designation | Extraction conditions | Number of channels | | Step, mm | | Material amount, mg/cm2 | Desing |
|---|---|---|---|---|---|---|---|
| | | X | Y | X | Y | | |
| P24 | FE, SRE, CSE | 16 | 16 | 2 | 1,5 | 175 | Ni capillaries |
| P26 | FE, SRE, CSE | 16 | 16 | 2,5 | 1,5 | 175 | Ni capillaries |
| P30 | FE, SRE, CSE | 16 | 16 | 1,6 | 1,6 | 140 | Ni capillaries |
| P1/chan.8 | FE, SRE, CSE | 16 | 16 | 2,5 | 2,5 | 7 | Kapton |
| P2/chan.8 | FE, SRE, CSE | 16 | 16 | 2,5 | 2,5 | 7 | Kapton |
| IM/chan.8 | SRE, CSE | 16 | 16 | 5 | 5 | 360 | Kapton/Ioniz |
| P1/chan.22 | FE, SRE, CSE | 16 | 16 | 2,5 | 2,5 | 7 | Kapton |
| P2/chan.22 | FE, SRE, CSE | 16 | 16 | 2,5 | 2,5 | 7 | Kapton |
| P3/chan.22 | FE, SRE, CSE | 16 | 16 | 2,5 | 2,5 | 7 | Kapton |
| P4/chan.22 | FE, SRE, CSE | 16 | 16 | 1 | 1 | 7 | Kapton |
| IM/chan.22 | SRE, CSE | 16 | 16 | 2,5 | 2,5 | 360 | Kapton/Ioniz |

Several radiation monitors (RM) along the extraction path make it possible to measure the beam losses in space and time at the partitions of deflecting magnets or on vacuum chamber walls.

When optimizing extraction with bent crystals, losses are measured using two independent systems integrated into the U-70 control system. Ionization chambers with air filling are used in these systems as sensors. One system continuously monitors losses over the entire U-70 ring. Another an independent system has sixteen monitors are placed in the extraction path and head the regions of the transport channels. The system provides multiple measurements of losses over the extraction time with a resolution of 10 ms.

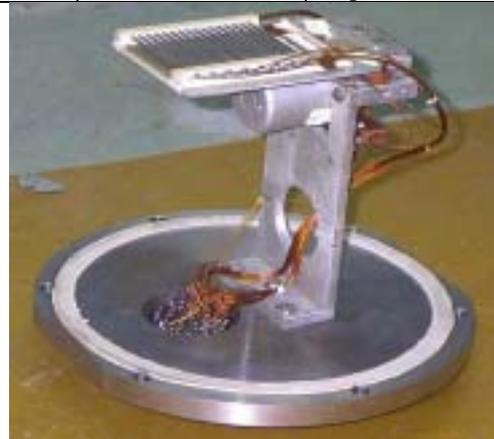

Figure11: A capillary profilometer with drive.

To make the proton beam guidance uniform, and to measure the relative intensity of the proton beam spilled onto crystals, feedback systems based on scintillation detectors were developed. They consist of two monitors each, installed at the orbit level and placed 6-20 m from the crystals.

The monitors include FEU-93 photomultipliers with (p- terphenyl) scintillators. Their radiation resistance means that they can be operated for two years without appreciable deterioration of their parameters. One monitor is the feedback monitor (FM); the other is for measuring relative intensity.

The signal from the relative intensity monitor is proportional to a beam fraction, which initiates nuclear reactions in the crystal material. The signal is lowest at an angular crystal position, and this feature is used when tuning the extraction conditions.

The set of diagnostic devices considered above, is not complete. When tuning the channeling conditions and optimizing the extraction channel, both a TV system and scintillation detectors are used.

The main advantages of such systems are simple implementation, a high threshold sensitivity, and a wide intensity range.

At the IHEP, this system is used to control the beam extraction at the U-70 accelerator. It includes 12 observation points. Six of these are arranged immediately in the U-70 ring facility and allow beam observation at positions of extraction elements.

## 5. CONDITIONS OF PROTON BEAM EXTRACTION WITH BENT CRYSTALS.

Proton beam extraction was studied mostly at energies of 70 and 50 GeV using strip-shaped (111) crystals (S-type) and (110) O-crystals (O-type). The crystal characteristics and the highest attained values of the extraction efficiency are listed in Table 4. During the experiment, the accelerated beam intensity was $(0.5-5)*10^{12}$ protons per pulse.

When the crystal began to operate in the channeling regime, its position was varied over angle. Figure 12 shows the typical orientation curve for 70-GeV protons extracted according to scheme 106-24-26 using an S-crystal no.1 (see Table 4). The operating position of the crystal corresponds to the curve maximum.

The extraction efficiency (the ratio of the extracted beam intensity to that of the intensity fraction guided to the crystal) was determined by the acquisition of statistical data during several hundred cycles. The measured and calculated (using available codes) extraction efficiencies in relation to the intensity fraction guided to the crystal at an energy of 70 GeV are shown in Fig.13, for extraction schemes 106-24-26 (Fig.13a) and 22-24-26 (Fig.13b).

Table 4

| № | Position | Type | Bending angle, mrad | Sizes l×h×R, mm | Efficiency, % | Energy, GeV | Extraction scheme |
|---|---|---|---|---|---|---|---|
| 1 | SS-106 | S | 1,0 | 2,0×35×0,5 | 85<br>80 | 70<br>70 | 106-24-26<br>106-20-22-26 |
| 2 | SS-106 | O | 0,7 | 3,5×5,0×0,7 | 60 | 70 | 106-24-26 |
| 3 | SS-19 | S | 2,0 | 5,0×45×0,5 | 67 | 70 | 20-22-26 |
| 4 | SS-19 | O | 2,1 | 5,0×5,0×0,7 | 65 | 70 | 20-22-26 |
| 5 | Unit22 | S | 0,8 | 1,9×45×0,5 | 85 | 70 | 24-26 |
| 6 | Unit 22 | S | 0,9 | 1,8×45×0,5 | 80 | 50 | 24-26 |

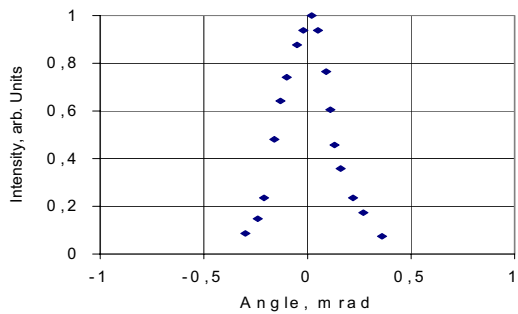

Fig.12. Extracted beam intensity as a function of the crystal orientation

For extraction according to scheme 106-24-26, the measured efficiency reaching ~85% decreases by ~10% as the beam fraction guided to the crystal increases. This is caused by the guidance angle drift (see Sec.3) resulting in a non-optimal angular orientation of the crystal with respect to the beam. The calculated curve (dashed line) is 3-5% higher than the experimental one, which indicates their satisfactory agreement.

For extraction according to scheme 22-24-26 (Fig.16b), a certain increase (~4%) in the efficiency is observed as the beam fraction guided to the crystal increases, This can be explained by assuming that the beam of unchanneled particles x partially arrives at the septum of the septum magnet in the first moment, hence, a certain beam fraction is lost. As the beam emittance decreases during extraction, these loses are lowered and the efficiency increases. A simulation of this process shoes that the experimentally observed dependence agrees well with the calculated one when a certain fraction of the unchanneled beam is incident onto a septum edge 0.3 mm deep (dashed curve 2). If a crystal position is changed, making it closer to the equilibrium orbit by ≥0.3 mm, the observed effect disappears, and the extraction efficiency increases by up to 90% (dashed curve 3). Thus, although we attained a high extraction efficiency of ~85%, calculations shoe that 90% is possible.

We also measured other characteristics of the extracted beam which are important for designing and performing physical experiments: its sizes, intensity, extraction duration, and operation stability.

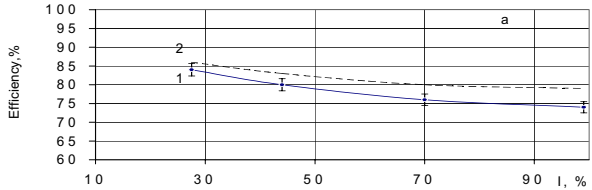
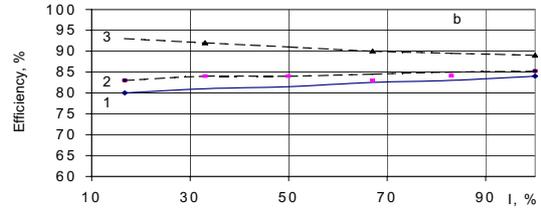

Figure 13: The extraction efficiency vs. the beam fraction guided onto the crystal for the extraction schemes (a) 106-24-26 and (b) 22-24-26.

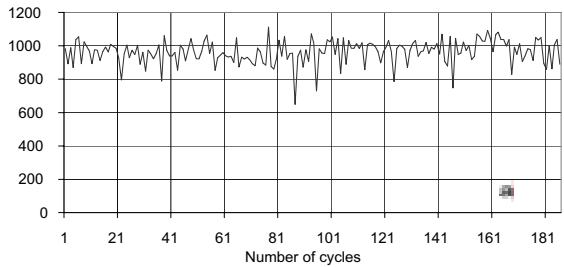
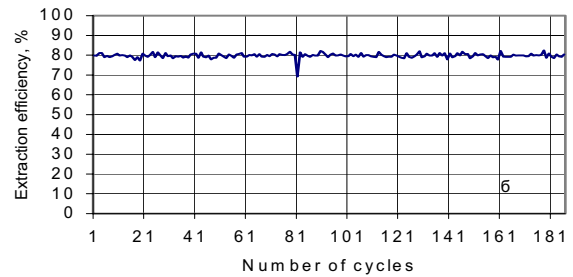

Fig. 14 Diagrams of (a) the intensity and (b) the efficiency of the 50 GeV proton beam extraction according to scheme 22-24-26.

Figure 14 illustrate reasonably stable operation of the extraction system in a certain time interval of 180 operation cycles of the accelerator. The extraction efficiency at an energy of 50 GeV reaches 80% and the extracted beam intensity is $10^{12}$ protons per pulse.

An important problem when applying crystals in accelerators is the radiation resistance. The critical particle flux, which the crystal can withstand, was experimentally estimated at the CERN and BNL as $\sim 2*10^{20}$ protons per sm2. Our experience showed that a crystal with an efficiency of 80-85% reliably allows the extraction of beams with an intensity of up to $10^{12}$ particles per pulse and a duration of 1-2 seconds per pulse, over more than two runs for 1400 hours each.

## 6. SIMULTANEOUS OPERATION OF PROTON BEAM EXTRACTION USING A BENT CRYSTAL AND TWO INTERNAL TARGETS.

Proton beam extraction using bent crystals allows the simultaneous operation of several internal targets, which was experimentally confirmed at the U-70 accelerator. Implementation of this mode opens up possibilities for the simultaneous operation of several experimental setups on the entire flattop of the accelerator magnetic cycle, which significantly reduce the costs of experiments. We should note that the classical resonant slow extraction is not compatible with the simultaneous operation of internal targets.

To implement the mode where the extraction operation with crystal and the internal targets operation are performed simultaneously, an extraction scheme with a crystalline station placed in straight section no. 19 (see Fig.4, Section I) was used. The proton beam was extracted using silicon crystals 5 mm long with a bending angle of from 1.7 to 2.3 mrad. Their efficiency reached ~60%. To guide the accelerated beam to the crystal and two internal targets T24 and T27 positioned, respectively, in the 24th and 27th magnetic units of the accelerator, it was necessary to cause a local orbit distortion of a particular shape (curve1 in Fig.4; the extracted beam path is curve 2). The system of local orbit distortion included three pairs of magnetic units from the U-70 accelerator. The beam guidance to the crystal and internal targets, as well as the maintenance of the regularity of the process, was provided by three beam feedback systems.

Implementing the extraction operation using a bent crystal simultaneously with internal targets required an increase in the number of control tools for the operation of extraction systems. One of the problems is that it is impossible to measure partial intensities during the simultaneous operation of several users. Furthermore, if the intensity consumed by the internal target is estimated using the so-called integrated monitors, whose signal is defined by the nuclear interaction of the proton beam with the target's material, this technique is inapplicable to the crystal. If the beam is extracted by the crystal, this signal depends not only on the spilled intensity, but also on the crystal orientation with respect to the incident beam, i.e., on the extraction efficiency.

To reliably measure the crystal operation efficiency under such conditions, the minimum set of monitored parameters should be as follows:

(i) the circulating beam intensity before extraction;
(ii) the total consumed intensity, or the difference between the accelerated intensity and the residue;
(iii) the intensity extracted by the crystal in the head region the transport channel and in the 30th straight section;
(iv) the signal of all the integrated monitors;
(v) the photomultiplier high voltages providing operation of the beam feedback systems.

Examples of the results from these measurements are shown in Figs.16, 17, and 18.

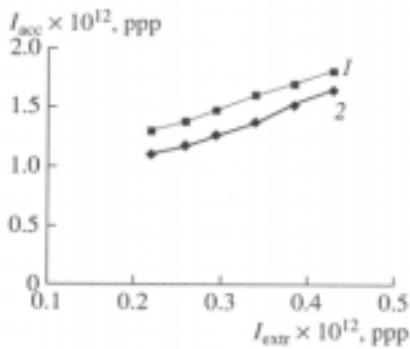

Fig. 15. The dependence of the accelerated beam intensity (1) and the consumed beam intensity (2) on the intensity extracted by the crystal.

It is impossible to directly measure the crystal efficiency under conditions of simultaneous operation with the internal targets. Therefore, to estimate the efficiency under these circumstances, one should use information from the integrated monitors employed in the beam steering onto internal targets. By changing the extracted intensity fraction while retaining the conditions of beam guidance to the targets, one can evaluate the crystal's operation efficiency. An illustration to this is the experimentally measured dependences of the accelerated beam intensity, as well as the total consumed beam intensity, on the beam intensity extracted by the crystal are shown in Fig.15.

The difference between the accelerated and consumed intensity is explained by the fact that it is expedient to have a certain (up to 20%) reserve of intensity to maintain long-term stable operation for the users. During this experiment, we gradually increased the beam intensity extracted by the crystal. We found that beam intensity consumed by the internal targets was not changed. Thus, the overall increase in the consumed intensity corresponded to the crystal operation. By determining the ratio of the extracted intensity increment to that of the consumed intensity increment from this curve, we obtain a crystal operation efficiency, in this experiment, equal to ~45%. In this case, it was unchanged as the extracted intensity increased twofold.

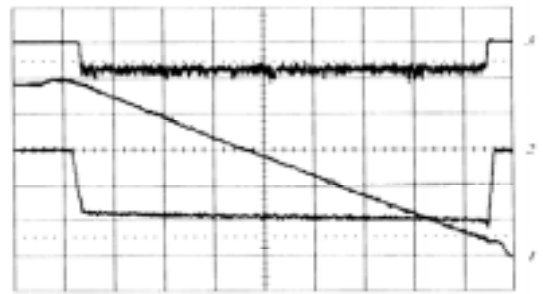

Fig. 16. Oscillograms of extraction conditions: (3) the feedback monitor signal, (2) the current in the system of local orbit distortion, and (1) the accelerated beam intensity. The scale divisions are $5*10^{11}$ protons (1) 50 A (2), 5V (3) in the vertical and 200 ms in the horisontal.

The efficiency of using the extracted proton beam in the experimental setup, in many respects, depends on the stability of the beam position on the external target and the beam size stability. As the experiment showed, in the case under consideration, the major causes of these parameters are changes in the current cycle of all the systems maintaining the beam guidance to the crystal and internal targets, as well as the instability of these currents from cycle to cycle. The current oscillogram in one of the guidance systems is shown in Fig. 16 (beam 2), as well as the intensity signal of the beam circulating in the accelerator (beam 1), and the signal of the crystal feedback monitor (beam 3).

The main operational characteristics for physical setups during the first three runs are listed in Table 5.

Table 5

| Main operation parameters | First run | Second run | Third run |
|---|---|---|---|
| Average intensity of the accelerated beam, protons per pulse | $3.0*10^{12}$ | $2.5*10^{12}$ | $2.7*10^{12}$ |
| Average used intensity, protons per pulse | $2.8*10^{12}$ | $2.2*10^{12}$ | $2.5*10^{12}$ |
| Bending angle of the crystal, mrad | 1.7 | 2.3 | 2.3 |
| Average intensity extracted by the crystal, protons per pulse | $3.0*10^{11}$ | $4.5*10^{11}$ | $5.5*10^{11}$ |

In these runs, the "Tagged Neutrino Facility" (TVF) positioned in channel 23 operated using the beam extracted by crystal (see Fig.5). To make this system operate, a high intensity of the extracted beam with the longest duration was required. Such requirements could not be met previously unless the extraction occurred using crystal. Table 5 shows that the average beam intensity extracted to the setup increased from run to run. It is imported to the efficiency of the use of the accelerated beam more than doubled.

The TNF was used to study three-particle decays of K± mesons: K+ →π +π 0π0, K-→π -π0π0. Since the K+ meson yield is significantly higher than the K- meson yield, one should proportionally increase or decrease the extracted proton beam intensity when changing the conditions. To provide such conditions, a technique for changing the operating voltage of the feedback monitor responsible for beam guidance to the crystal was used (see Fig.17). We can clearly see from this figure that there is good agreement between the change in the secondary particle intensity (I23) in the experimental setup and readings (S19) of the integrated monitor displaying the proton beam flux onto the crystal. In this case, readings (S24, S27) of integrated monitors used in the guidance to targets were virtually unchanged.

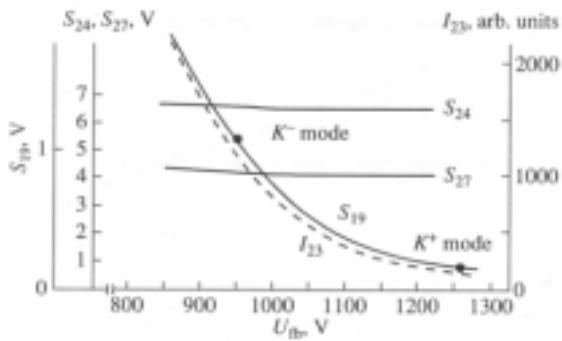

Figure 17: The parameters of the simultaneous (parallel) mode vs. the photomultiplier voltage in the feedback system providing guidance onto the crystal. S19, S24 and S27 are reading of integrated monitors, I23 is the secondary particle intensity in the experimental setup.

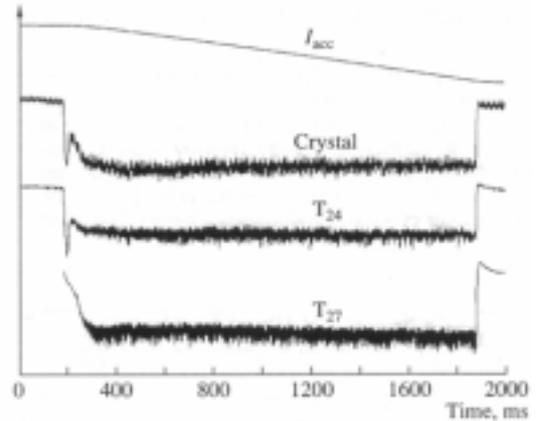

Figure 18: Signal oscillograms of the monitors providing beam guidance to the crystal and the interval targets T24, T27.

One of the most important parameters of extracted beams is the time structure (or the effective extraction time). In the mode under consideration, the time structure is convenient for all the simultaneously operating experimental setups. Some current oscillograms of corresponding monitors are shown in Fig.18.

The effective extraction time during such an operation reached 98%. The characteristics of the produced beams satisfy experimenters for the following set of parameters: the beam intensity, duration, quality, and sizes on targets. As an example Figure 22 shows the profile on the target of the vertex detector spectrometer (VDS) positioned in channel no.19 (see Fig.5).

The proton beam extraction scheme using bent crystals, developed and implemented at the U-70 accelerator, offers methods that can be used for further modifications. In particular, in order to increase the number of experiments simultaneously carried out at the U-70 accelerator, one more crystalline deflector station with a crystal bend at an angle of ~9 mrad (corresponding to the direction of channel no.22) was mounted in straight section no.30. This crystal deflected a small fraction (~$10^7$ protons per pulse) of the beam, already extracted toward channel no.23, to another experimental setup positioned in channel no.22 (see Fig.5).

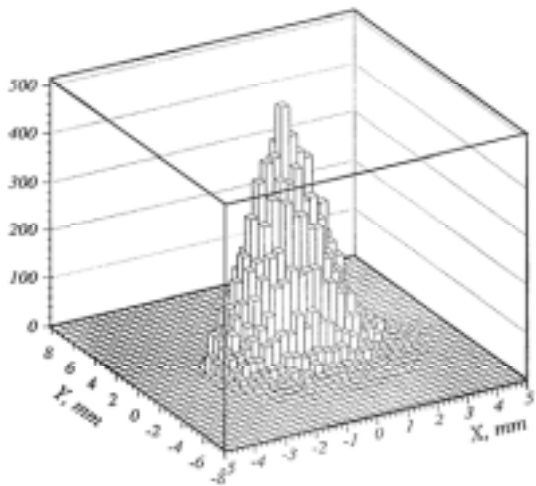

Figure 19: Channeled beam profiles in the VDS experimental setup. X and Y are the horizontal and vertical planes, respectively.

Due to such conditions, it became possible to increase the number of simultaneously operating setups on the flattop of the magnetic cycle by up to five, without taking into account the possibility of operating two more internal targets in "shadow", without feed back.

In the long term, when operating with a 50 GeV proton beam, it becomes possible to increase the duration of the flattop of the magnetic cycle by up to 4 s. For this purpose, the use of the developed extraction schemes using bent crystals seems to be most adequate and productive.

Since 1999, the use of bent crystals, to maintain the high-energy physics program, became systematic in all the runs of the U-70 accelerator.

## 7. TESTBED FOR STUDYNG CRYSTALS.

As it was mentioned earlier (see p.6 and fig.4) in the test region the station Si84 and Si86 are placed to test crystals and to carry out the experiments with crystals in beam collimation mode. This testbed was organised near the beam loss localisation system of the
U-70 acceleratior.

The loss localization system of the U-70 accelerator includes a system of local orbit distortion, the absorber itself, and a set of devices for diagnosing and controlling the beam parameters. The layout of the testbed is shown in Fig.20.

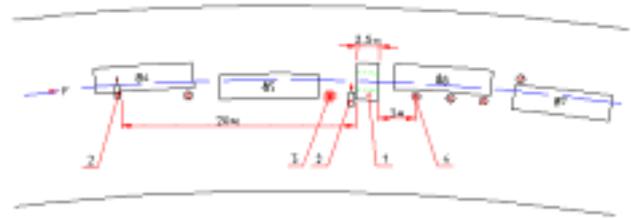

Figure 20: The layout of crystalline stations and beam diagnostic tools near the absorber: (1) absorber, (2) crystalline stations, (3) profilometer, (4) ionization chambers, and magnetic units nos. 84-87,

The system of local orbit distortion allows orbit deformation in the region where the absorber is placed. The beam displacement velocity is controlled by the front of the bumps and is ~3 mm/ms. The absorber position and the shape of local orbit distortion are shown in Fig.21. Here the bottom numerals correspond to the indices of the accelerator's magnetic units, and the vertical scale is the distance from the equilibrium orbit.

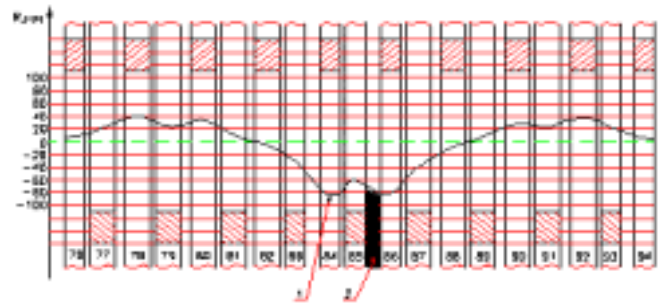

Figure 21: Schematic diagram of beam guidance to the absorber: (1) the shape of the local orbit distortion and (2) the beam absorber.

Several stations with crystalline deflectors were installed in the 84th and 86th sections.
During studies with crystal deflectors placed in the 84th and 86th sections it was also necessary to measure such parameters as the beam position, size, intensity, and radiation field near the absorber. To this end, a complex system for measuring the beam parameters was developed. To measure the radiation field of secondary particles, eight loss monitors, based on ionization chambers filled with air, were mounted near the absorber. These chambers were positioned at a distance of ~600 mm from the vacuum chamber of the accelerator. Moreover, a scintillation detector was used to control the beam guidance onto the absorber and crystal. Thus a testbed was created for a wide range of studies with crystals in the entire energy range of accelerated protons from 1.5 to 70 GeV.

The absorber positioned near the circulating beam results in the fact that secondary particles escaping from it land on the accelerator equipment. Therefore, the absorber should be characterized by parameters where the dumping of protons onto it does not cause the irradiation of magnetic units above an acceptable value controlled by the number of protons reaching the absorber and absorbed in it. A steel absorber, 2.5 m long, installed at the U-70 decreases the total energy of the beam of dumped 70GeV protons by ten times.
Such an attenuation is attained at a sufficient displacement of the proton beam from the absorber edge. Since particles are displaced by several micrometers per revolution during the elimination of the beam halo, or during the interception of lost protons using bumps, nuclear interactions of protons take place on the absorber edge. In this circumstance, particles deflected due to scattering into the absorber are absorbed quite efficiently; however, particles deflected in the opposite direction are not absorbed. When operating under these conditions, the energy absorption efficiency is ~50% . As the beam deflection from the absorber edge increases, the efficiency rises and reaches ~90% in the case of delivery at 20mm from the edge. The absorber thickness chosen in the radial direction is 40 mm. The radial and vertical absorber sizes are dictated by the vacuum chamber aperture.
The crystal operation efficiency was determined using profilometers. The quality of the collimation system's operation was estimated using ionization chambers placed near the collimator, as well as a system for beam loss measurement over the entire accelerator perimeter. Efficient beam absorption in the collimator depends heavily on the depth of the beam dump at the collimator entrance face. Therefore, beam location using a crystal requires optimization of parameters such as the bending angle and the coordinate with respect to the collimator boundary. The efficiency of the crystal itself is undoubtedly a determining factor.

The efficiency of the crystalline deflectors was measured by calibrating the profilometers using kicker magnet of the extraction system. In this case, the entire beam was kicked far beyond the absorber edge and the total profilometer signal corresponded to 100% of the beam intensity. If the beam is kicked by the crystal, the ratio of the total signal to that of the signal obtained using the kicker magnet yields the magnitude of the efficiency of the crystal being studied. In the case under consideration, the determination of the accuracy of this parameter depends on the accuracy of the measured signals. An analysis of the measurement carried out showed that this total error does not exceed 5%.
The same tested was used in experiments at a beam guidance velocity of ~0.7 mm/ms to a crystal with a bending angle of 0.8 mrad and a length of 1.7 mm along the beam. Figure 22 shows the results of these measurements, i.e., the beam density distributions in the horizontal plane formed by employing the kicking magnet and crystal. It follows from these measurements that the efficiency reaches 85%; that is, the same value as during the beam extraction from the accelerator (see Sec.6).

The dependences of changes in the signals measured using the ionization chamber, when the crystal was displaced in misaligned and aligned states, are shown in Fig.23. We can see that putting the crystal into the channeling mode results in a significant decrease in the number of secondary particles detected by the chamber. This is due to a decrease in the number of nuclear interactions of protons with the crystal.

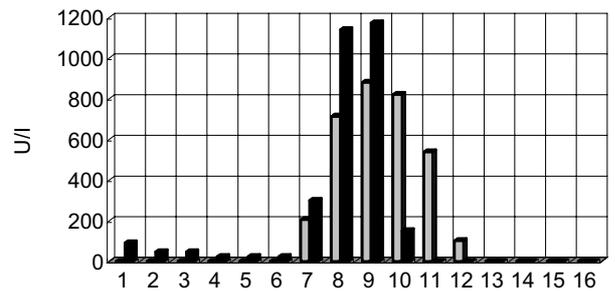

Fig.22. Beam profiles formed at the absorber by the kicker magnet (light area) and aligned crystal (dark)

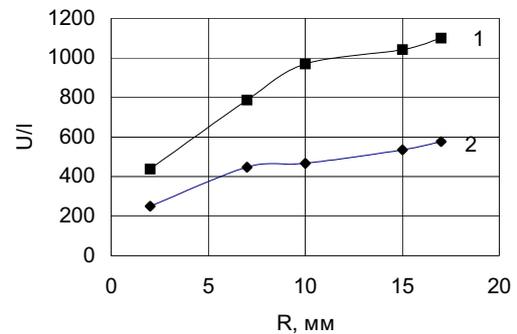

Figure 23: The ionization chamber signal vs. the coordinate of (1) misaligned and (2)aligned crystals. The signal amplitude is normalized to the intensity.

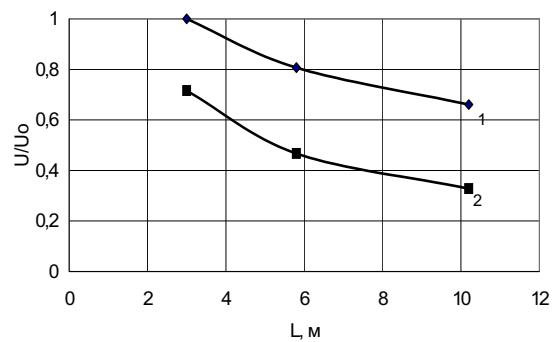

Figure 24: Changes in ionization chamber signals when the beam is dumped onto the absorber (1) using bumps and (2) the aligned crystal (SS84) with the efficiency ε=60%. L is the distance from the absorber in meters.

The major factor confirming the efficiency of the application of crystalline deflectors to collimate the beam is the experimental estimates of the total radiation field near the absorber. To this end, we carried out measurements with ionization chambers placed behind the absorber, as is shown in Fig.20. The radiation field depends on the ratio of the fraction of particles incident into the material depth to those on its edge. Figure 24 shows changes in the radiation fields behind the absorber under various operating conditions of the loss localization system. The value $\frac{u}{u_0} = 1$ corresponds to the reading of the first ionization chamber behind the absorber in the case of dumping by a bump.

We can see from Fig.24 that when crystals with an efficiency of ~60% are used, the readings from the ionization chambers near the 86th unit show a significant decrease in the radiation field in comparison with the beam guidance to the absorber edge using bumps. However, fluxes of secondary particles, generated by the interaction of protons with the absorber, irradiate equipment not only near the absorber but elsewhere as well. Irradiation of the U-70 accelerator equipment was additionally monitored by taking measurements of the secondary particle flux using the loss measurement system. We showed that when operating with an aligned crystal, these fluxes are weaker by a factor of 1.6 than when the beam is localized using bumps. These data are in agreement with the measurement results shown in Fig.24. It is clear that radiation fields in the accelerator will further decrease as the crystal efficiency increases.

Arrangement of crystalline deflector stations near the absorber allowed experiments a wide energy range. The results of IHEP accelerator experiments on the determination of the crystal efficiency at various proton energies, as well as the numerical simulation results, are shown in Fig.25. A crystal 1.8*27*0.5 mm$^3$ in size, with a deflection angle of ~0.8 mrad, was used in these experiments.

Figure 25 a shows good agreement between the measured and calculated efficiency of the bent crystal. The efficiency degradation as the proton energy decreases is mainly explained by increasing the rootmean-square angle of the multiple Coulomb scattering and decreasing the dechanneling length. The dependence obtained also shows that the use of the same crystal allows beam extraction in an acceptably wide energy range of 40-70 GeV with an efficiency of higher than 60%.

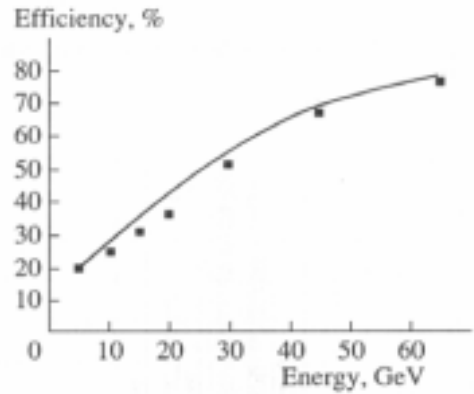

Figure 25: The proton-energy dependence of the crystal efficiency: (solid curve) calculation and (squares) experiment.

We also note that the dependences shown in Fig.25 were measured using crystals with parameters optimized to attain the extraction efficiency for the 70GeV particle beam. Therefore, if we extrapolate the calculated curve to energies above 70 GeV, it reaches a maximum at 70 GeV and begins to decay rapidly due to an increase (proportional to the particle energy) in the critical radius of the crystal.

## CONCLUSIONS

The studies carried out at the IHEP accelerator showed that the use of short silicon crystals ~2 mm long makes it possible to attain a high efficiency of beam extraction, ~85%.

The extraction using short bent crystals, developed at the IHEP accelerator, significantly extends the capability of experiments involving high-energy beams. This provides, depending on experimental requirements, an intensity of extracted beams in the range of $10^6$-$10^{12}$ particles per pulse, thus allowing simultaneous operation of several experimental setups to collect data and reducing the time of experiments. This extraction is a good addition to the resonant slow extraction existing at the IHEP accelerator, which provided proton beam intensities of $5*10^{11}$-$10^{13}$ particles per pulse. More details can be found in our publications [1-28].

It would be expedient to make this extraction competitive with the resonant slow extraction; i.e., to increase the intensity of extracted beams by ten times (from $10^{12}$ to $10^{13}$ protons per pulse). It seems that this problem can be solved using more heat-and radiation-resistant crystals instead of silicon ones, e.g., artificial diamond, whose parameters are higher by factors 3 and 6, respectively.

We believe that the high-efficiency extraction of accelerated particles using short bent crystals, developed at the IHEP accelerator, will stimulate further development of this method at other high-energy accelerators.